\documentclass[twocolumn,bibnotes,showpacs]{revtex4-1}

\usepackage{graphicx}
\usepackage{color}
\usepackage{soul} 
\usepackage{amsmath}
\usepackage{hyperref}
\usepackage{braket}
\hypersetup{colorlinks=true, citecolor=cyan, urlcolor=blue, linkcolor=blue}

\newcommand{\PRL}[4]{\textit{#1}, Phys. Rev. Lett. \textbf{#2}, #3 (#4)}
\newcommand{\PRA}[4]{\textit{#1}, Phys. Rev. A \textbf{#2}, #3 (#4)}
\newcommand{\PRB}[4]{\textit{#1}, Phys. Rev. B \textbf{#2}, #3 (#4)}
\newcommand{\PRX}[4]{\textit{#1}, Phys. Rev. X \textbf{#2}, #3 (#4)}
\newcommand{\PRApplied}[4]{\textit{#1}, Phys. Rev. Appl. \textbf{#2}, #3 (#4)}
\newcommand{\Science}[4]{\textit{#1}, Science \textbf{#2}, #3 (#4)}
\newcommand{\Nature}[4]{\textit{#1}, Nature \textbf{#2}, #3 (#4)}

\begin{document}

\title{Weakly-invasive probing of a superconducting resonator based on electromagnetically induced transparency}
\title{Tuneable and weakly-invasive probing of a superconducting resonator based on electromagnetically induced transparency}

\author{Byoung-moo Ann}
\email{byoungmoo.ann@gmail.com}
\author{Gary. A. Steele}
\affiliation{Kavli Institute of Nanoscience, Delft University of Technology, 
2628 CJ Delft, The Netherlands} 
\date{\today}

\begin{abstract}
Superconducting cavities with high quality factors play an essential role in circuit quantum electrodynamics and quantum computing. 
In measurements of the the intrinsic loss rates of high frequency modes, it can be challenging to design an appropriate coupling to the measurement circuit in such a way that the resulting signal is sufficiently strong but also that this coupling does not lead to unwanted loading circuit, obscuring the intrinsic internal loss rates. 
Here, we propose and demonstrate a spectroscopic probe of high-Q resonators based on the phenomena of electromagnetically-induced transparency (EIT) between the resonator and qubit in the weak dispersive coupling regime. 
Applying a sideband drive signal to the qubit, we observe an interference dip originated from EIT in the qubit spectroscopy, originating from the quantum interference between the qubit probe signal and sideband transition.
From the width and the depth of the dip, we are able to extract the single-photon linewidth of the resonator from an analytical model.
Working in a previously unexplored regime in which the qubit has a larger linewidth than the resonator reduces the technical challenge of making a high-coherence qubit and is advantageous for remaining in the weakly-invasive limit of coupling to the resonator.
Furthermore, the sideband and the dispersive coupling between the resonator and the qubit can be tuned $in~situ$ controlling the strength of the sideband drive power. 
This $in-situ$ tuneability allows the technique to be applied for efficient measurement of the resonator loss rate for any quality factor below a fixed upper bound, on the order of $10^8$ for our device, allowing a wide range of quality factors to probed using a single design.


%

\end{abstract}

\maketitle
\section{Introduction}

Superconducting resonators with high quality factors play an important role in the fields of quantum science and information. One example includes quantum computing based on encoding of quantum information in the bosonic modes represented superconducting resonators, which is particularly attracive for the implementation of quantum error correction\cite{Bosonic-0,Bosonic-01}. In this scheme, a harmonic system such as an ion's mechanical mode or a photonic mode carries the quantum information instead of a two-level quantum bit \cite{Bosonic-1,Bosonic-2}. For these, and many other, applications, a longer lifetime of the resonator is highly desired. However, in order to implement near-lossless resonators, it is crucial to have a technique that is able to determine what the intrinsic linewidth is of the superconducting resonator at the single photon level.


A common spectroscopy approach for determining the loss rate of a superconducting resonator is to couple the resonator to external transmission lines and measure its transmission or reflection spectrum. A disadvantage of this approach is that the resonator loss rate induced by the external channel ($\kappa_e$) can dominate the total loss rate $\kappa$ of the resonator. 
In principle, internal and external loss rates ($\kappa_{i,e}$) can still be independently extracted, although in practice imperfections in the microwave impedance of the measurement setup can result in asymmetric lineshapes that complicate the independent determination of the two\cite{Fano, Fano-2}. In particular, extracting  $\kappa_i$ can become challenging when $\kappa_e \gg \kappa_i$. 
One approach for mitigating this problem is to ensure that $\kappa_e \ll \kappa_i$, in which case the internal loss rate is determined by the total linewidth\cite{High-Q}. A disadvantage of this approach, however, is that the signal-to-noise ratio of the measurement is reduced, and it can become challenging to measure at single-photon excitation levels.

Determining the appropriate value of $\kappa_e$ to design for such spectroscopy requires $a~priori$ estimate of the order of magnitude of $\kappa_i$, which presents a challenge as the value of $\kappa_i$ can be difficult to predict. This is a problem in particular in  planar resonators, in which $\kappa_i$ can be strongly affected by surface contamination that is difficult to control.
An alternative non-invasive technique, which is unaffected by the possible unknown impedances of external circuitry and in which the coupling to the resonator can be tuned in-situ, could be advantageous for spectroscopy of superconducting cavities.

Here, we present a weakly-invasive spectroscopic probe of a superconducting resonator using a qubit in the weak dispersive coupling limit. The underlying principle is based with electrically induced transparency (EIT) \cite{eit-1,eit-2,eit-3,Long-PRL2018,eit-4, Novikov-NPhys2016, Liu-PRA2016}. When the qubit has a broader linewidth than that of the resonator, if the sideband coupling is smaller than the difference between them (EIT regime), we can find a narrow dip in the qubit population spectrum, which is a strong indication of EIT. From these measurements, The qubit and resonator decay rate can be extracted independently using a model of the EIT process. As long as the sideband coupling can be made the same order of magnitude as the resonator linewidth, one can extract the decay rate of the resonator in a way that is insensitive to the error in the qubit decay rate.

The EIT-based spectroscopy approach is weakly invasive in the sense that it is sufficient that the qubit and the resonator are in the weak dispersive coupling, in which there is very little direct hybridisation between the two.
Consequently, loss rate through the qubit ($\kappa_{q}$) can be chosen to be negligible. 
In the presence of a sideband drive of appropriate frequency, the drive induces a predictable and tunable hybridisation between the qubit and resonator through sideband transitions, which is the technique we apply here to perform resonator spectroscopy.
Furthermore, being based on inducing a small EIT window in the qubit spectrum, our approach does not require high coherence qubit, and wide ranges of values of $\kappa_{i}$ of the resonator can be accurately probed in spectroscopy by tuning the sideband drive power.

We experimentally demonstrate this scheme with experimental observations using a device based on a transmon qubit \cite{Koch} coupled to a coplaner waveguide (CPW) resonator. The work presented here demonstrates a application of EIT related physics in a weak dispersive coupling regime. EIT with a circuit quantum electrodynamics (QED) platform has already been reported in several configurations \cite{Long-PRL2018,eit-4, Novikov-NPhys2016, Liu-PRA2016} but the present paper is the first study of EIT related phenomena with a dispersively coupled resonator and a qubit system when the qubit has a broader linewidth.

This paper is organized as follows. In Sec.~\ref{2}, we discuss theoretical background of this work. We present experimental results in Sec.~\ref{3}. We provide  analysis of the results, discussion on a major error source and a further direction of the study in Sec.~\ref{4}. A summary of this work is given in Sec.~\ref{5}.

\section{Theoretical description}
\label{2}
\subsection{Qubit population spectrum under sideband driving}
\label{2-1}

\begin{figure}[htbp]
\centering\includegraphics[width=0.48\textwidth]{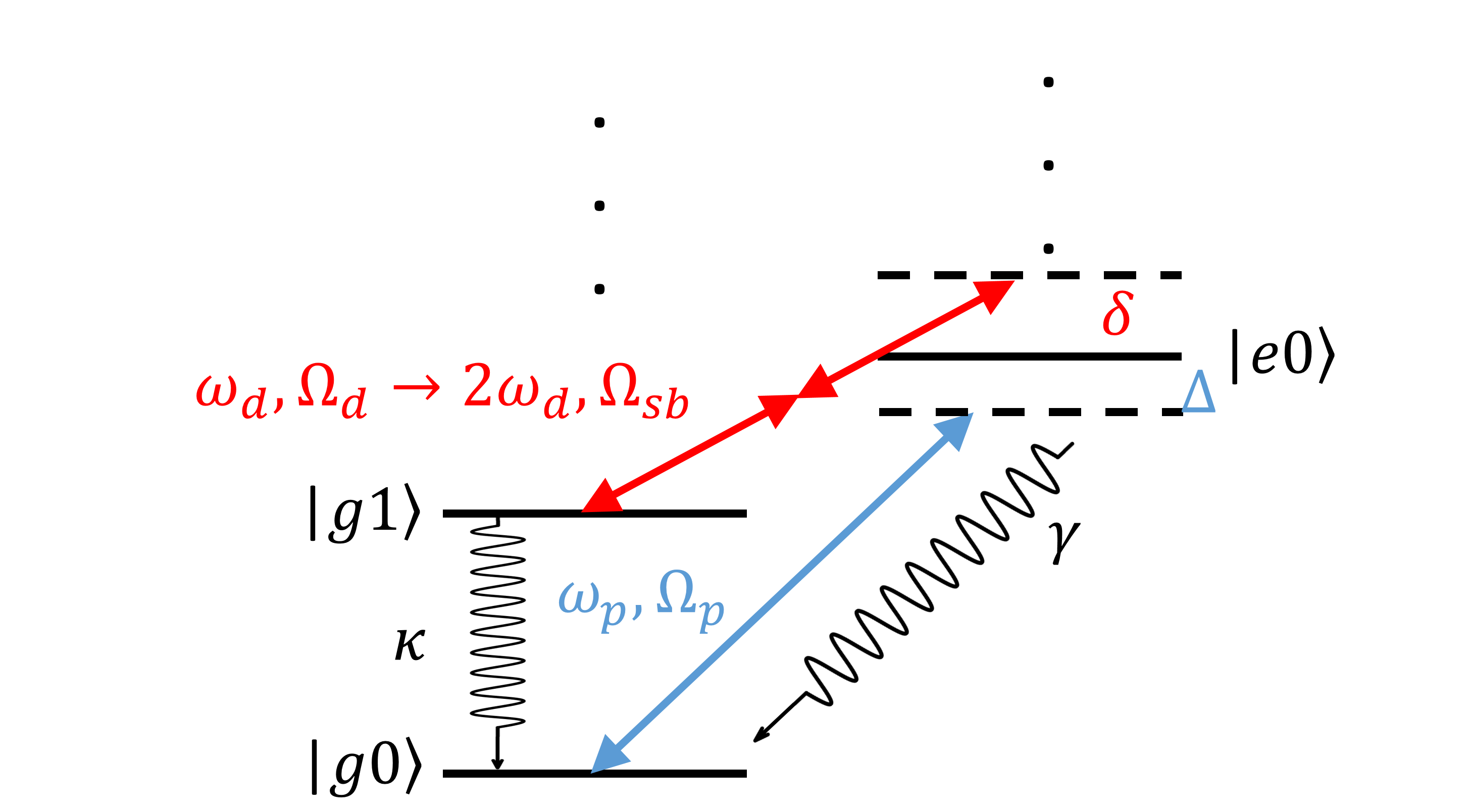}
\caption{Energy level diagram describing the the system of a dispersively coupled target resonator and qubit. Red arrows indicated an off-resonant two-photon drive field for enabling two-photon sideband transitions. The blue arrow indicates a weak resonant field for probing the qubit. The readout resonator is omitted in the diagram.}
\label{Schematic_thy_1}
\end{figure}

The system studied is depicted in Fig.~\ref{Schematic_thy_1}. A two-level qubit is dispersively coupled to a target resonator whose internal loss rate is our interest. 
We induce a first order sideband transition with a coupling rate of $\Omega_{sb}$ that couples $\ket{g,n}$ and $\ket{e,n-1}$ (red arrows) through an external coherent drive with frequency $\omega_d$. The first order sideband transition is dipole forbidden for transmon qubits as described in \cite{Blais}. Therefore, a two-photon process is used to enable this transition in this work.

In the figure, $\ket{g}$ and $\ket{e}$ refer to the ground and excited state of the qubit. The numbers refer to the photon number of the target resonator.
The target resonator frequency and the Stark-shifted transition frequency of the qubit are $\omega_t$ and $\omega_q$ respectively. Their decay rates are given by $\kappa$ and $\gamma$ respectively. A weak field (blue arrow) probes the qubit transition. We define the detuning between qubit and probe $\omega_q-\omega_p$ to be $\Delta$. Also, we define $\delta$ as $\omega_q-\omega_t-2\omega_d$. These definition are also graphically presented in Fig.~\ref{Schematic_thy_1}.
In the real device, we additionally have a readout resonator coupled dispersively to the qubit, which is not present in Fig.~\ref{Schematic_thy_1}. We will also omit the readout resonator in the following mathematical derivation since it has no role in featuring the EIT.

The effective Hamiltonian of the system in Fig.~\ref{Schematic_thy_1} is
\begin{equation}
\label{eq: red sideband fitting model}
\begin{split}
    \hat{\mathcal{H}} = \,& \frac{{\omega}_q}{2} \hat{\sigma}_z + {\omega_t} \hat{a}^\dagger \hat{a} - 2\chi_{qc} \hat{\sigma}_z \hat{a}^\dagger \hat{a} \\
    & + \frac{\Omega_{sb}}{2} \left( \hat{a} \hat{\sigma}_{+} e^{-2i\omega_d t} + \hat{a}^\dagger \hat{\sigma}_{-} e^{+2i\omega_d t} \right) \\
    & + \frac{\Omega_p}{2} \left( \hat{\sigma}_{+} e^{-i\omega_p t} + \hat{\sigma}_{-} e^{+i\omega_p t} \right),
\end{split}
\end{equation}
where $\hat{\sigma}_z$ denotes the Hamiltonian of the two-level qubit, $\hat{\sigma}_{\pm }$ are the raising and lowering operators of the qubit state, and $2\chi_{qt}$ is the dispersive shift between the qubit and the resonator.
The external drive results in a negligible change of the of the dispersive shift \cite{Blais} and is therefore neglected.
By applying the following time-dependent unitary transform, 

\begin{equation}
    \hat{\mathcal{U}} = \exp \left[ i(\omega_p -2\omega_d )t \hat{a}^\dagger \hat{a} + i(\omega_p)t \hat{\sigma}_z \right].
\end{equation}
the Hamiltonian can be simplified to

\begin{equation}
\label{eq: red sideband fitting model2}
    \begin{split}
    \hat{\mathcal{H'}} = \,& \frac{{(\omega}_q- \omega_p)}{2} \hat{\sigma}_z + {(\omega_c+ 2\omega_d - \omega_p)} \hat{a}^\dagger \hat{a} - 2\chi_{qt} \hat{\sigma}_z \hat{a}^\dagger \hat{a} \\
    & + \frac{\Omega_{sb}}{2} \left( \hat{a} \hat{\sigma}_{+} + \hat{a}^\dagger \hat{\sigma}_{-}  \right) \\
    & + \frac{\Omega_p}{2} \left( \hat{\sigma}_{+} + \hat{\sigma}_{-} \right).
\end{split}
\end{equation}
Here, we can use the definitions of $\Delta=\omega_q- \omega_p$ and  $\delta=\omega_q- \omega_t - 2\omega_d$; Both $\Delta$ and $\delta$ are as defined in Fig.~\ref{Schematic_thy_1} to simplify the expression. 
Then, the Hamiltonian takes the form
 
\begin{equation}
\label{eq: red sideband fitting model3}
    \begin{split}
    \hat{\mathcal{H'}} = \,& \frac{\Delta}{2} \hat{\sigma}_{z} + (\Delta-\delta) \hat{a}^\dagger \hat{a} - 2\chi_{qt} \hat{\sigma}_z \hat{a}^\dagger \hat{a} \\
    & + \frac{\Omega_{sb}}{2} \left( \hat{a} \hat{\sigma}_{+} + \hat{a}^\dagger \hat{\sigma}_{-}  \right) \\
    & + \frac{\Omega_p}{2} \left( \hat{\sigma}_{+} + \hat{\sigma}_{-} \right).
\end{split}
\end{equation}
The dynamics of the system is then given by the Lindbald equation:
\begin{equation}
\label{eq: red sideband fitting model4}
	\begin{split}
   	\frac{d\hat{\rho}}{dt} =  \vphantom{\sum_n} -\frac{i}{\hbar} \left[ \hat{\mathcal{H'}}(t), \hat{\rho}(t) \right]\\
	+\frac{\gamma}{2}\mathcal{D}[\hat{\sigma}_{-}]\rho +  \frac{\kappa}{2}\mathcal{D}[\hat{a}]\rho+\frac{\gamma_{\phi}}{2}\mathcal{D}[\hat{a^{\dagger}}\hat{a}]\rho,
\end{split}
\end{equation}
where $\mathcal{D}[\mathcal{\hat{O}}]\hat{\rho} = 2\mathcal{\hat{O}}\hat{\rho}\mathcal{\hat{O}}^\dagger - \mathcal{\hat{O}}^\dagger\mathcal{\hat{O}}\hat{\rho}-\hat{\rho}\mathcal{\hat{O}}^\dagger\mathcal{\hat{O}}$,
$\kappa$ is the decay rate of the target resonator, and $\gamma$ is that of the qubit. $\gamma_{\phi}$ is the pure dephasing rate of qubit.
From the steady state solution $\hat{\rho}_{ss}$ that satisfies $d\hat{\rho}_{ss}/dt$=0, one can obtain the steady state qubit population by tracing out the resonator state, ${\rho}_{ee}=\textup{Tr}_{res}[\hat{\rho}_{ss}(1+\hat{\sigma}_z)/2]$.

\subsection{Resonator spectroscopy at single-photon levels using sideband transitions}
\label{2-2}

\begin{figure}[htbp]
\centering\includegraphics[width=0.48\textwidth]{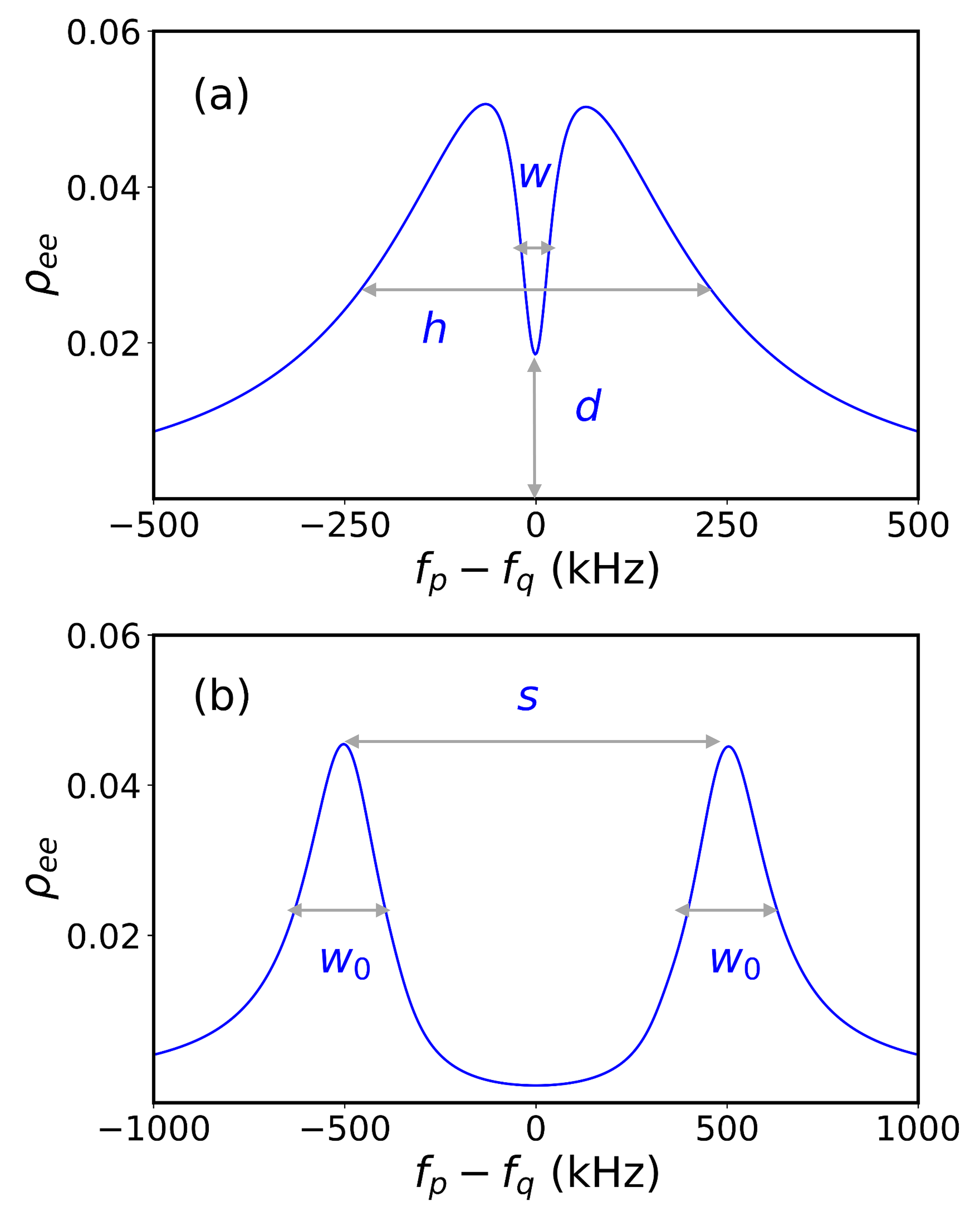}
\caption{Single-photon resonator spectroscopy through sideband transition. $\delta$ is set to 0 for both the EIT and ATS regime simulations. 
(a) Simulated qubit population spectrum in EIT regime when the parameters $\Omega_{sb},\Omega_{p},\gamma,\gamma_{\phi}, \kappa,\chi_{qt}$ are 2$\pi\times$(100,100,400,0,30,10) kHz (solid line). As an indication of the electromagnetically induced transparency, a Lorentizan dip (interference dip) appears in the qubit population spectrum, which is characterized by its width $w$ and minimum population $d$. 
(b) Simulated qubit population spectrum in ATS regime when the parameters $\Omega_{sb}, \Omega_{p},\gamma,\gamma_{\phi},\kappa,\chi_{qt}$ are 2$\pi\times$(1000,100,400,0,30,10) kHz (solid line). In this regime, the spectrum has two peaks separated by $s$ and their widths are both $w_0$.
In the simulation, we set $\gamma_{\phi}=0$, typically condition for the fixed frequency transmon qubit used in our experiment. 
In Sec.~\ref{A0}, we present additional simulation results that show how the $\gamma_{\phi}$ changes the qubit population spectrum. 
}
\label{Schematic_thy_2}
\end{figure}

When $\left | \gamma+2\gamma_{\phi}-\kappa \right | >\Omega_{sb}$  and $ \gamma+2\gamma_{\phi} > \kappa$, the sideband transition leads to a narrow transparency window in the qubit transmission spectrum.
This results from the interference between two different transitions, $\ket{g,n+1} \rightarrow  \ket{e,n}$ and $\ket{g,n} \rightarrow  \ket{e,n}$.
In this work, we measure the population of qubits $\rho_{ee}$ rather than the transmission $\textup{Im}[\rho_{eg}]$.
We define the qubit spectrum as the response of its average population as a function of the probe frequency $\rho_{ee}(\omega_p)$.
In Fig.~\ref{Schematic_thy_2}(a), we simulate the qubit population spectrum with reasonable parameters satisfying the EIT condition based on the master equation in Section~\ref{2-1}.
We can also find the same features of the transmission spectrum in the qubit population spectrum as well.
A Lorentzian dip in the qubit population spectrum, in the following we will refer to this as an `interference dip', is characterized by its width ($w$) and minimum population ($d$).

In the linear response limit ($\Omega_{p}  \ll  \Omega_{sb}$) and for zero detuning ($\delta = 0$), the width and depth of the dip are given by $w\sim\gamma+2\gamma_{\phi}+\kappa-\sqrt{{(\gamma+2\gamma_{\phi}+\kappa)}^{2}-\Omega_{sb}^{2}}$, $d\sim{\omega_p}/({\gamma+2\gamma_{\phi}+\frac{\Omega_{sb}^{2}}{\kappa^2}})$ and the total linewidth is $h \sim \gamma+2\gamma_{\phi}$ \cite{Liu-PRA2016}.
When $\Omega_{p}$ and $\chi_{qt}$ are known, $\kappa$,$\gamma$,$\gamma_{\phi}$ and $\Omega_{sb}$ together characterize the population spectrum and thus one can extract these by fitting the spectrum to the model calculated by the master equation in Section~\ref{2-1}.

Too small a value of $\Omega_{p}$ requires an excessive measurement time.
Fortunately, although $\Omega_{p}  \ll  \Omega_{sb}$ does not hold in the present simulation in Fig.~\ref{Schematic_thy_2}(a), we can clearly see the dip in the spectrum unless the $\Omega_{p}$ is excessively large.  Nevertheless, the upper bound on $\Omega_{p}$ sets a limit to the feasibility of our approach for investigating a single-photon level resonator loss rate. 
In the experiment, $\Omega_{p}$ is separately calibrated as described in Appendix~\ref{A3}. Also, $\chi_{qt}$ can be calculated from the device parameters.
In this work, since the qubit is weakly coupled to the resonator, the calculated $\chi_{qt}$ is only $2\pi \times$7.8 kHz.
With this magnitude, it hardly affects the spectrum and we confirm that neglecting $\chi_{qt}$ does not make a significant difference in the fitting results.


In order to achieve the EIT condition, it is possible to achieve a sufficient sideband coupling rate $\Omega_{sb}$ for spectroscopy even with a very weak dispersive coupling between the qubit and the resonator. To be able to extract the intrinsic resonator linewidth accurately from the fits, one needs to be in the regime of $\Omega_{sb} \sim \kappa$. By using a strong sideband drive strength, this can be achieved in a limit where the dispersive coupling to the qubit results in a negligible modification of the resonator linewidth. 

Specifically, the contribution of the dispersive coupling to the qubit to the resonator loss rate, which we denote as $\kappa_{q}$, scales with $(g_{qt}/\Delta_{qt})^2$, where $g_{qt}$ and $\Delta_{qt}$ are bare coupling rate and detuning between the qubit and the target resonator respectively. By arranging a large detuning between the qubit and the target resonator, this can be made negligbly small. The sideband coupling $\Omega_{sb}$, however, scales as $g_{qt}(\Omega_d/\Delta_{qd})^2$, where $\Delta_{qd}$ is $\omega_{q}-\omega_{d}$.
The large $\Delta_{qd}$ can be compensated for by a large sideband drive strength. In this way, the EIT spectroscopy technique can be made minimally invasive on the resonator it is probing. This is in contrast, for example to other qubit-based spectroscopy approaches \cite{PNS} that requires a strong dispersive coupling. We also note that while the sideband drive strength is strong, it is highly off-resonant from the resonator itself and the occupation number of the resonator remains negligible.

In Fig.~\ref{Schematic_thy_2}(b), we also simulate the qubit population spectrum when the Auther—Towns splitting \cite{Novikov-PRB2013} condition $\left | \gamma+2\gamma_{\phi}-\kappa \right | < \Omega_{sb}$ holds.
Unlike EIT, ATS arises from the result of electromagnetic pumping that results in a dressed normal mode splitting of the two modes in the rotating frame of the pump.
In circuit QED platform, ATS is also widely explored in several configurations \cite{Mika-PRL2009, Kelly-PRL2010,Li-SciRep,Novikov-PRB2013}.
When $\delta=0$, the spectrum has two symmetric peaks separated by $\Omega_{sb}$ and each linewidth is equal to $(\gamma+2\gamma_{\phi}+\kappa)/2$. 
Both the qubit and resonator decay equally characterize the linewidth of each peak.
Thus, one cannot set both $\kappa$ and $\gamma$ as free fitting parameters and thus the qubit decay rate should be separately calibrated.
This can be more problematic if an extremely small decay rate of resonator is expected.
In this case, the measured resonator decay rate becomes sensitive to the error in the qubit decay rate measurement, unless the qubit decay rate is much smaller than the resonator's decay rate, which imposes a challenging requirement on the preparation of the device. 


\section{Experiment}
\label{3}
\subsection{Device configuration}
\label{3-1}

\begin{figure}[htbp]
\centering\includegraphics[width=0.48\textwidth]{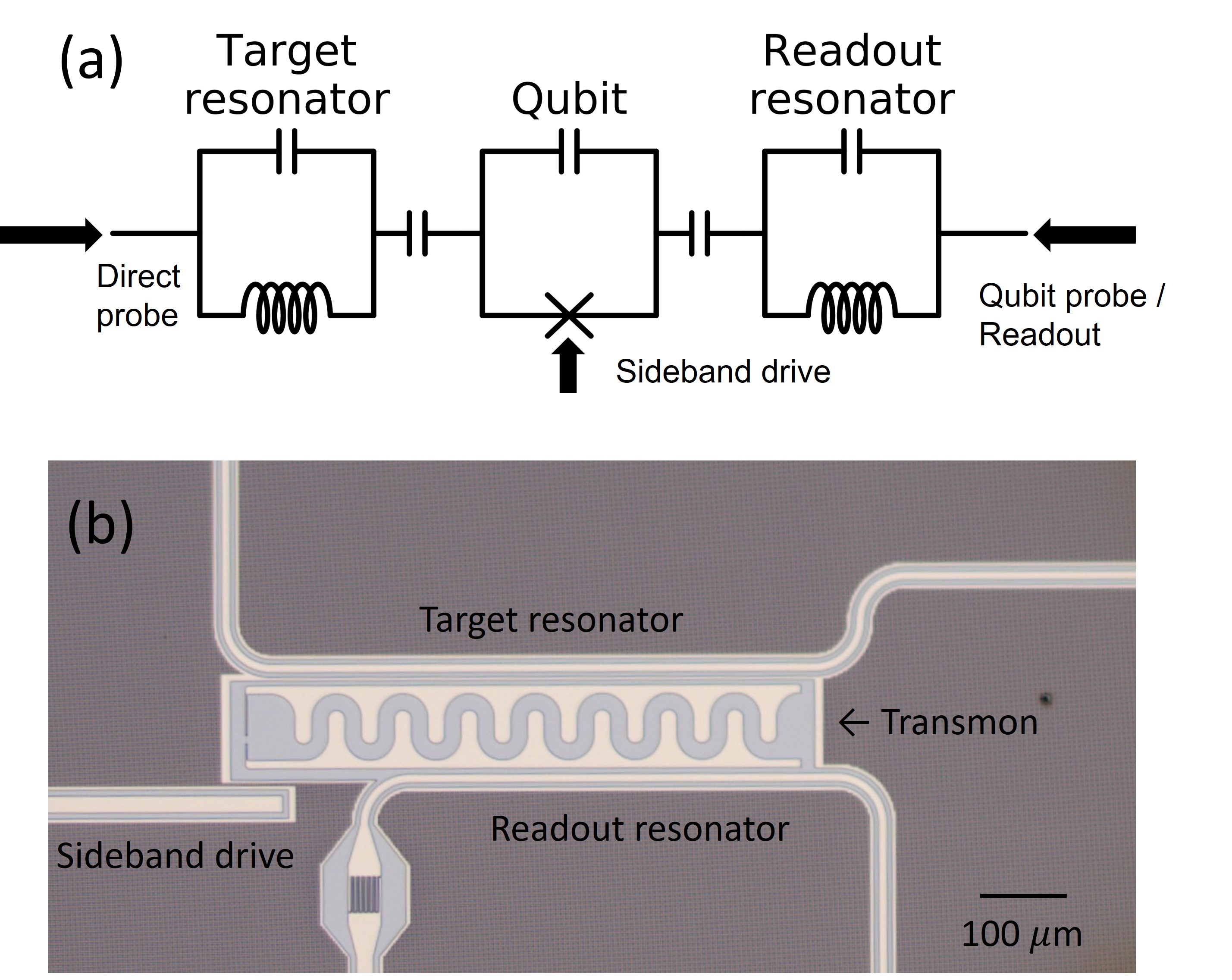}
\caption{Experimental setup. 
(a) Simplified circuit diagram of the device used in the experiment.
(b) Optical microscopy image of the superconducting circuit. }
\label{Schematic_exp}
\end{figure}
Fig.~\ref{Schematic_exp}(a) presents a simplified circuit diagram of the device used in the experiment.
More detailed information on the circuit design and related electronics can be found in Appendix~\ref{A1}.
A transmon qubit ($\omega_{q0}/2\pi$=6.723 GHz, without sideband drive) is capacitively coupled to two $\lambda/4$ co-planar waveguide (CPW) resonators. One is the target resonator ($\omega_{t}/2\pi$=2.9 GHz). The other is the readout resonator ($\omega_{r}/2\pi$=4.07 GHz) to measure the qubit population more efficiently. Both are dispersively coupled to the qubit with dispersive coupling $\chi_{qt}/2\pi$=7.8 kHz and $\chi_{qr}/2\pi$=1.3 MHz to the target and readout resonators respectively. The bare coupling between each resonator and qubit is estimated by $g_{qt}/2\pi=58$ MHz and  $g_{qr}/2\pi = 193$ MHz respectively. Qubit decay rate $\gamma$ is around $2\pi \times$400 kHz. Each resonator is inductively coupled to different feedlines. These values yield qubit limited resonator decay rate $\kappa_{q}/2\pi \approx 100$Hz, which is far below than typically achieved internal loss rate in our lab (around $2\pi\times 5-20$ kHz.) 

\begin{figure}[htbp]
\centering\includegraphics[width=0.48\textwidth]{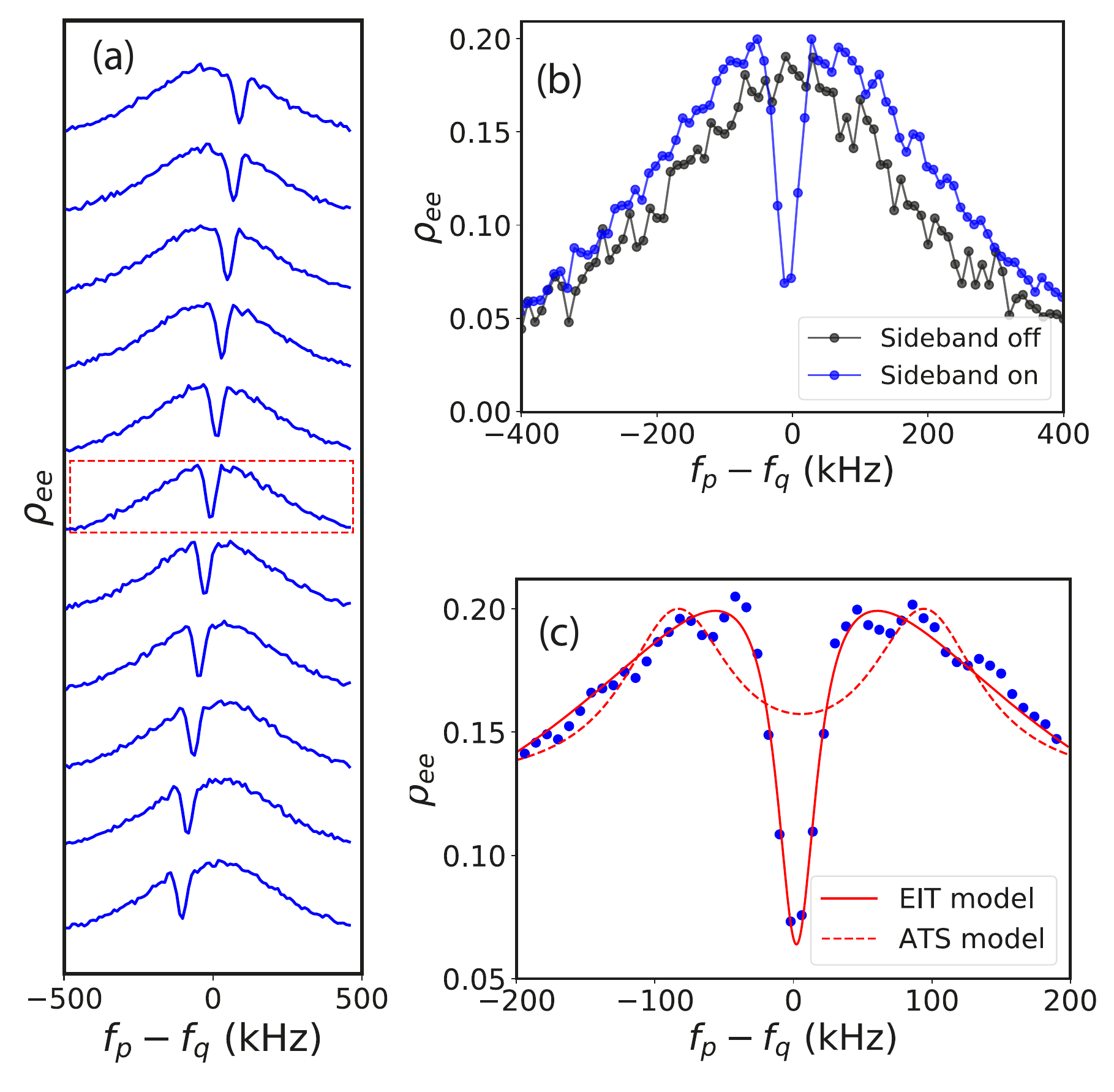}
\caption{Observation of interference dip in a qubit population spectrum. (a) Observation of interference dip in qubit population spectrum while scanning $\delta$ from -200 to 200 kHz. The spectrum when $\delta=0$ is enclosed by a red box. (b) Comparison between qubit population spectrum with (blue) and without (black) sideband field. (c) A separate measurement near the interference dip with finer step. The validity of the EIT model is confirmed while the ATS model breaks down.}
\label{exp1}
\end{figure}

We use a single-junction fixed frequency qubit and therefore its transition frequency is insensitive to magnetic flux noise, a strong source of dephasing in flux tuneable transmon qubits. This is also experimentally confirmed by the observation of $T_2\simeq 2T_1$ in separate time domain measurements given in Appendix~\ref{A2}. Although it is not necessary in our case, it could be advantageous to use a flux tuneable qubit, with which $\kappa_q$ is $in~situ$ tuneable by adjusting the detuning to the ressonator.  
The technique and analysis present here is also applicable for the flux tuneable qubit. We discuss how our scheme is extended for the flux tuneable qubit in Sec.~\ref{4-3}.

The optical microscopy image of the circuit can be found in Fig.~\ref{Schematic_exp}(b). The transmon qubit and CPW resonators were patterned on a 100 nm niobium titanium nitride (NbTiN) film on a Silicon substrate \cite{SRON}. The Josephson junction of the qubit is made by Al-AlOx-Al.

\subsection{Experimental results}
\label{3-2}
We apply the sideband drive directly to the Josephson junction of the circuit through the direct drive line (middle arrow in Fig.~\ref{Schematic_exp}(a)). The qubit probe $\Omega_{p}$ is applied through the feedline coupled to the readout resonator (right arrow). 
In addition, we have a direct probe of the target resonator (left blue arrow). We use in all four microwave sources in the experiment. One is used for the qubit probe, another for the qubit readout, another for the sideband drive, and the remaining one for the direct resonator probe. In order to avoid measurement induced broadening in the qubit population spectrum, we performed the measurement in pulsed configuration. First, a 20-$\mu$s long probe pulse is applied, rapidly followed by a 200-ns long readout pulse. Using this pulsed readout scheme, the qubit population spectrum is unaffected by the photons in the readout resonator during  qubit measurement.

Fig.~\ref{exp1}(a) shows the measured interference dip in the qubit population spectrum, which results from a sideband transition between the qubit and the target resonator. The sideband drive frequency $\omega_d$ is swept around $2\omega_d = {\omega}_{q} - \omega_c$ and we find $\delta\approx 0$ when $\omega_{d}/2\pi$=1.945 45 GHz.  The interference dip is conspicuously identified in the comparison to the spectrum without sideband transition in Fig.~\ref{exp1}(b). The probe amplitude $\Omega_p$ is $2\pi\times 264$ kHz according to the calibration method presented in Appendix~\ref{A3}. The value we chose is a compromise between the high contrast of the interference dip and the proper measurement time. The sideband drive also shifts the qubit frequency. In Appendix~\ref{A4}, we present the data for how much the qubit frequency is shifted when we obtain a sufficient sideband coupling.

\begin{figure}[htbp]
\centering\includegraphics[width=0.48\textwidth]{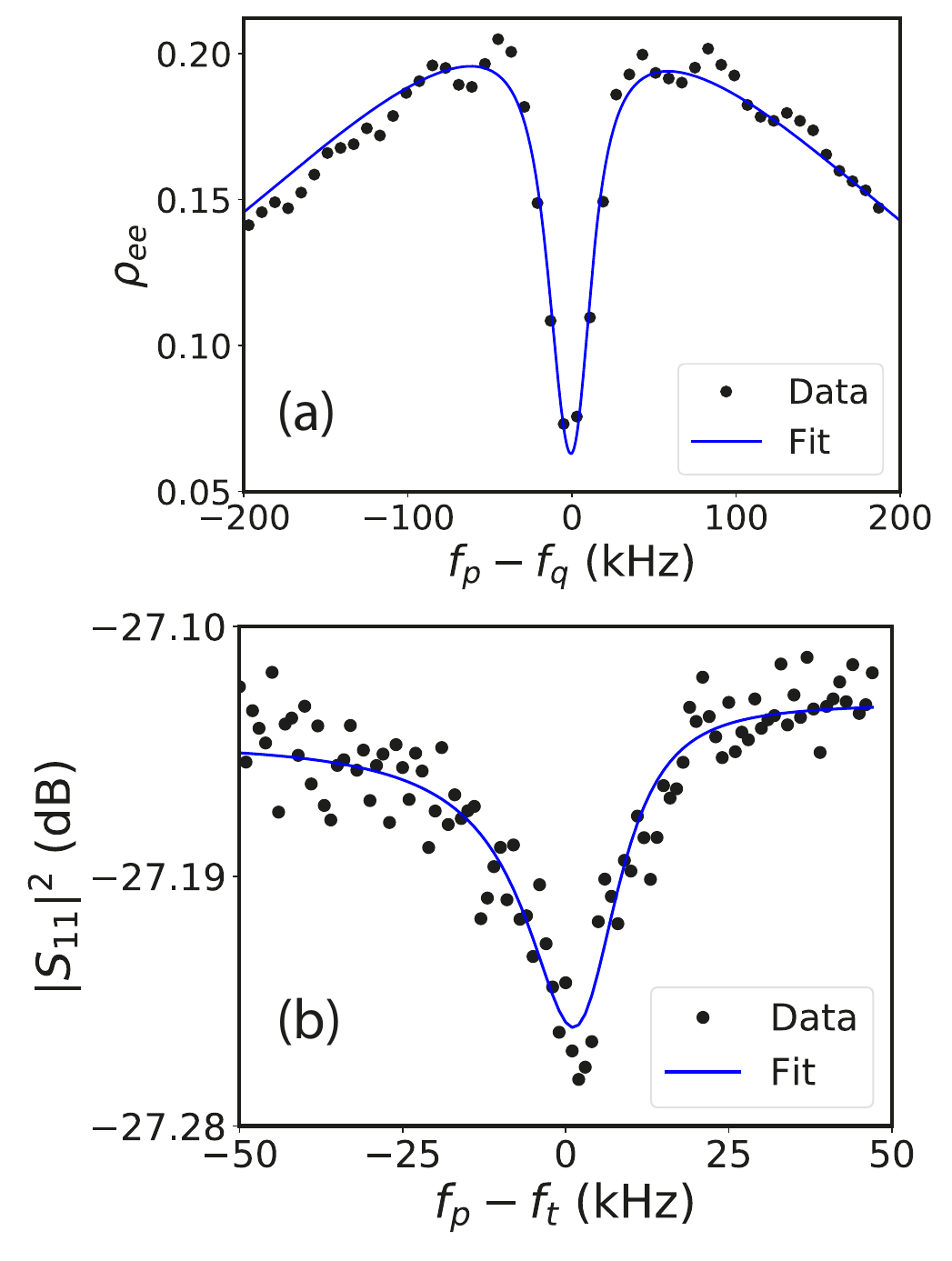}
\caption{Two different spectroscopic measurement of a superconducting resonator. (a) Spectroscopy of the superconducting resonator mode using interference dip. From numerical fitting of the qubit population spectrum calculated using the master equation, we extract a resonator linewidth of $\kappa/2\pi = 20.3 \pm 1.5$ kHz. (b) A direct reflection measurement of the resonator through the weakly coupled port. Even for very weak external coupling, the resonator displays an asymmetric lineshape due to Fano resonance. From a fit to a Fano resonance lineshape, we extract a linewidth $\kappa/2\pi = 17.2 \pm 1.8$ kHz, in agreement with the results of the spectroscopy based on interference dip within the experimental error.}
\label{exp2}
\end{figure}

In addition to the process of EIT, observations similar to those in Figure \ref{exp1} can also arise from the process of ATS. In order to distinguish EIT from ATS, one can numerically fit the data by given simplified model in linear response limit
\cite{Anisimov-PRL2011}.
A system in the EIT regime can be modeled by
\begin{equation}
\label{eq1}
\rho_{ee,\textup{EIT}}(\omega_p)=\frac{C_{+}^{2}}{\Delta^2+\gamma_{+}^{2}}-\frac{C_{-}^{2}}{\Delta^2+\gamma_{-}^{2}}
\end{equation}
When the system is in the ATS regime, $\rho_{ee}$ in the linear response limit is
\begin{equation}
\label{eq2}
\rho_{ee,\textup{ATS}}(\omega_p)=\frac{C^{2}}{(\Delta-\Delta_0)^2+\gamma_0}+\frac{C^{2}}{(\Delta+\Delta_0)^2+\gamma_0}.
\end{equation}
All the parameters in these expressions are free fitting parameters except $\Delta$. Here, $\gamma_{0,\pm}$ is not necessarily the same with $\gamma$ and $\gamma_{\phi}$.
We perform a numerical fit with the two different fitting models above corresponding to each phenomenon. The results are given in Fig.~\ref{exp1}(c). The data is taken with the same conditions as in (b) but a different range and step of probe frequencies. While the EIT model Eq.~\ref{eq1} shows excellent agreement with the data (solid line), the ATS model Eq.~\ref{eq2} fails to explain the data well. Accordingly, the fact that the system is in the EIT regime is clearly demonstrated. It is notable that the EIT model is still applicable to the data even when $\Omega_{p} \ll \Omega_{sb}$ does not hold.

We fit the qubit population spectrum with the numerical model in Section~\ref{2-1} to extract the target resonator's linewidth $\kappa$. The result is presented in Fig.~\ref{exp2}(a). The data is the same as in Fig.~\ref{exp1}(c). From the data, $\kappa/2\pi\approx20.3 \pm 1.5$ kHz, $\gamma/2\pi\approx445.95 \pm 2.4$ kHz and $\Omega_{sb}/2\pi=112\pm0.5$ kHz are extracted. When $f_p=f_q$, the photon number in the target resonator is approximately 0.3, based on the master equation solution with extracted parameters. The estimated $\kappa_{q}$ is only $2\pi\times$ 104 Hz.

We also measured the single-photon level $\kappa$ by the normal reflection spectrum via a weakly coupled port in Fig.~\ref{exp2}(b) to verify the above result. A Fano resonance is also considered in the fitting process. From the fitting, $\kappa/2\pi\approx17.2 \pm 1.8$ kHz is obtained. The upper bound of the resonator photon number is approximately 1.25 based on the input power from the source of the probe, the room temperature, and the cryogenic wiring. 
The $\kappa$'s extracted from both approaches agree within the overlapping statistical error.

\section{Discussion and outlook}
\label{4}
\subsection{Analysis of the results}
\label{4-0}

We measured the single-photon level $\kappa$ of the target resonator through two independent approaches. 
The target resonator is coupled to the external environment through the qubit ($\kappa_q$) and also the external feedline($\kappa_{e}$). Both coupling rates are $2\pi\times104$Hz and $2\pi\times112$Hz respectively based on the measurement.
Although these two quantities are similar, one can see significant difference in the contrast of the spectroscopic signal. 
In Fig.~\ref{exp2}(a), the suppression of the qubit population in the EIT based spectroscopy is more than 5 dB. On the contrary, the suppression of the reflection in Fig.~\ref{exp2}(b) is only 0.1 dB.
This clearly shows that our approach covers much wider range of $\kappa_i$.

For given design in our work, considering an error margin of 10\% on $\kappa_{q}$, our EIT based spectroscopy works nicely for the resonator with an internal quality factor ($Q_{i}$) up to $10^{8}$ and a resonance frequency of 10 GHz. Since it is difficult to obtain single-photon level $Q_i$ more than $10^{8}$ for planar resonators,
the design is already optimized for such types of device.

\subsection{Effect of qubit decay rate fluctuation on the measurement}
\label{4-1}
\begin{figure}[htbp]
\centering\includegraphics[width=0.48\textwidth]{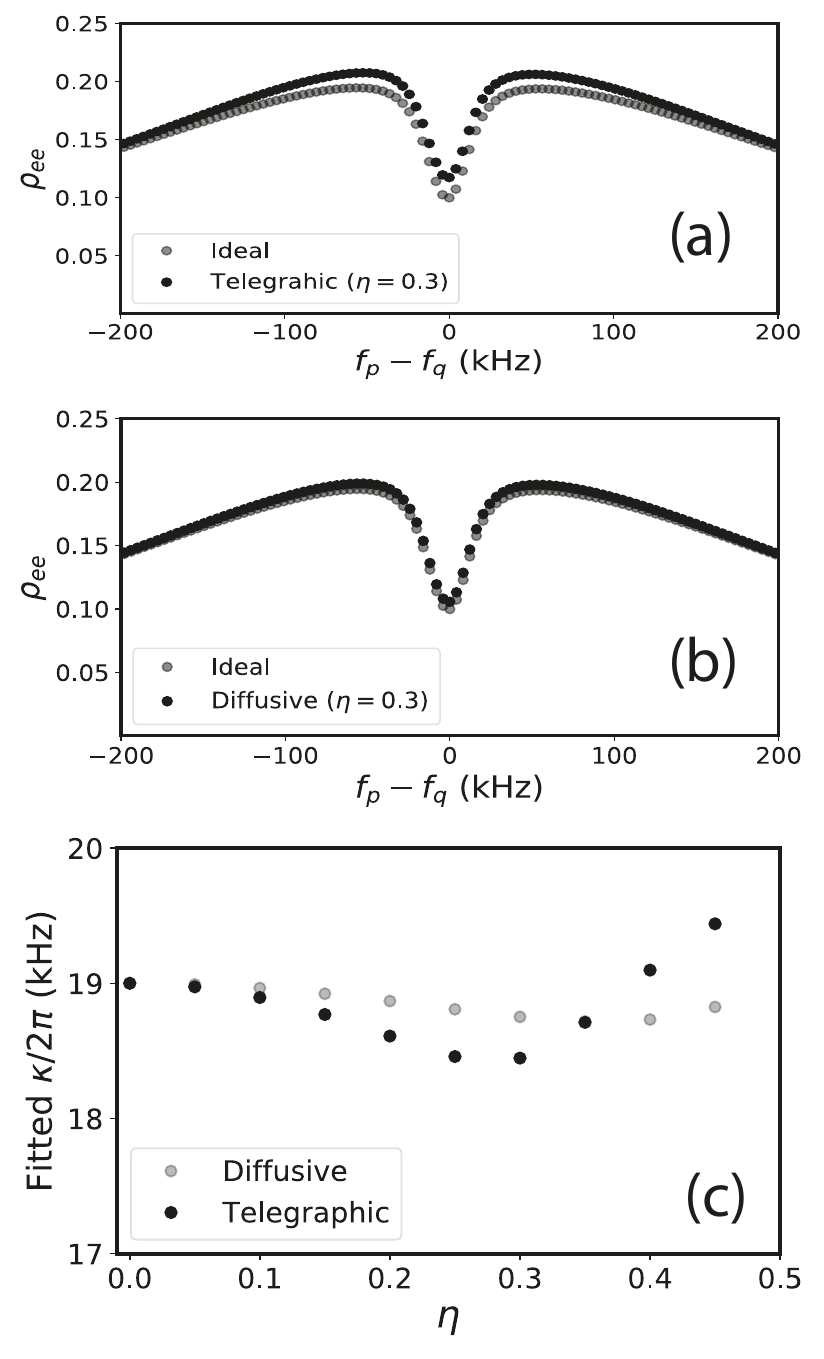}
\caption{Investigation of systematic error induced by fluctuation in qubit decay rate. See main text for the detailed method and a definition of $\eta$.
(a) and (b) How the qubit population spectrum varies when the qubit decay rate fluctuates in a telegraphic way (a) and a drifty way (b) when $\eta$ is 0.3 for both cases.  
(c) Effect of the qubit decay rate fluctuation on fitted $\kappa/2\pi$. Telegraphic fluctuation (gray) has a larger effect than a drifty fluctuation (black) for given $\eta$.   Mean qubit decay rate over measurement duration ($\gamma_0$) is $2\pi\times$450 kHz and $\Omega_{sb}, \Omega_{p}, \kappa, \chi_{qt}$ are 2$\pi\times$(100,264,19,7.8) kHz in the simulation.}
\label{Error}
\end{figure}

It is often observed that the decay time of a superconducting qubit can fluctuate in time \cite{Klimov-PRL2018}.
If one cannot finish the measurement before the fluctuation happens, there is a resulting distortion in the qubit population spectrum. 
In this subsection, we model such distortion and simulate how it affects the fitted $\kappa$ depending on the degree and the tendency of the fluctuation.

We assume that we rapidly sweep the frequencies, faster than the time scale of the fluctuation, but repeat the sweeping enough to obtain an adequate signal to noise ratio.
We consider two different trends in the fluctuation: telegraphic fluctuation and diffusive fluctuation\cite{Klimov-PRL2018}. In the simulation, the qubit decay rate varies from $\gamma_{i}=(1-\eta)\gamma_0$ to $\gamma_{f}=(1+\eta)\gamma_0$ during the measurement. For the telegraphic case, we assume the decay rate jumps at the middle of the measurement. For the diffusive case, the decay rate varies at a constant rate over time.  

We define $\rho_{ee}(\omega_p;\gamma)$ as the qubit population spectrum when the qubit decay rate is $\gamma$.
For telegraphic fluctuation, the spectrum is expressed by
\begin{equation}
\rho_{ee}^{tele}(\omega_p)=
\frac{1}{2}[\rho_{ee}(\omega_p;\gamma_i)+\rho_{ee}(\omega_p;\gamma_f)]. 
\end{equation}
For diffusive fluctuation from $\gamma_{i}$ to $\gamma_{f}$ homogenously, the spectrum is expressed by
\begin{equation}
\rho_{ee}^{diff}(\omega_p)=
\frac{1}{n}[\Sigma^{n}_{k}\rho_{ee}(\omega_p;\gamma_i+2\eta k/n)], 
\end{equation}
where $n$ is the number of sweeps during the measurement, set to 100 in our simulation here. 

The results of the simulation can be found in Fig.~\ref{Error}.
In Fig.~\ref{Error}(a) and (b), we compare the qubit population spectrums with and without qubit decay rate fluctuation for $\eta=0.3$. The gray curves indicate the spectrum when $\gamma$ is fixed at $\gamma_0$=$2\pi\times$450 kHz. The black curves indicate the spectrum distorted by fluctuations in $\gamma$. In Fig.~\ref{Error}(c), we fit the distorted spectrum with the ideal fitting model from Section~\ref{2-1} and how the extracted $\kappa$ is affected. 

\subsection{Further direction of the study}
\label{4-2}
In this work, we rely on a two-photon assisted transition, the achieved sideband coupling strength is only 0.1 percent of the bare coupling strength between the qubit and the resonator. Achieving larger couplings could be achieved by introducing other types of qubit, for example a flux qubit, with which one can address the first order sideband transition to the resonator with a single photon transition. In that case, the required bare coupling for the desired sideband coupling strength becomes significantly smaller, along with a smaller requirement for $\kappa_q$ to stay in the EIT regime.

The resonator spectroscopy scheme presented here is extensible to the case of many target resonators having different frequencies, as long as a qubit is coupled to them with the proper coupling strength.
Typically, the spectroscopy of multiple resonators on a chip requires a circuit design with a long feedline so that all the resonators are properly coupled to the feedline.
Such a structure could induce some slotline mode and limit the scalability of the design. For our method, such a long feedline is not necessary as one only needs to feed the probe and readout pulse to the qubit, providing a relatively simple measurement technique for the spectroscopy of multiple resonators on a chip.
For for this, an X-mon \cite{Xmon} or star-mon \cite{Starmon} design for the qubit, for example, would allow the single qubit to couple to multiple resonators.

\subsection{Extension to flux-tuneable qubit}
\label{4-3}
As we discuss above, employing a flux-tunenable qubit enables $in-situ$ control of $\kappa_q$. This is useful when we need control $in-situ$, which is not necessary in our work. 
In this case, unlike the case of the fixed frequency qubit, we need to take the pure dephasing rate of the qubit ($\gamma_{\phi}$) into account. 
This however does not add complexity in using our scheme. 
As long as we have $\kappa_q \ll \kappa_i$ and $\gamma\sim\gamma_{\phi}$, the effect of the pure dephasing of the qubit to the resonator is still negligible. 
In fitting process, we would only need to include $\gamma_{\phi}$ in the master equation model. The effect of the pure dephasing in the qubit population spectrum is distinguishable from other parameters and therefore we can sucessfully extract the $\kappa$ from the fitting even with non-zero $\gamma_{\phi}$.

\section{Conclusion}
\label{5}
To summarize, we have demonstrated a single-photon resonator spectroscopy using a weakly coupled qubit. From the appearance of an electromagnetically induced transparency in the qubit population spectrum, we obtained a single-photon linewidth of a high-Q resonator. We validated our result using an independent measurement of the resonator linewidth through a separate transmission line. Our spectroscopy method here is compatible with a resonator of an even smaller loss rate than that in the present work, without demanding a high coherence qubit, due to its being weakly invasive. This work offers a method for reliable estimates of the loss rates of superconducting resonators and enables the study of EIT in a weak dispersive regime of circuit QED.

\begin{acknowledgements}
We thank Wouter Kessels for his support in the data analysis.
We also thank David Theron and Jochem Baselmans for provide us with NbTiN film.
Byoung-moo Ann acknowledges support from the European Union’s Horizon 2020 research and innovation program under the Marie Sklodowska-Curie grant agreement No. 722923 (OMT).
This project has received funding from the European Union’s Horizon 2020 research and innovation programme under grant agreement No. 828826 - Quromorphic.
The data that support the findings of this study are available in \cite{data}.
\end{acknowledgements}

\begin{figure}[htbp]
\centering\includegraphics[width=0.48\textwidth]{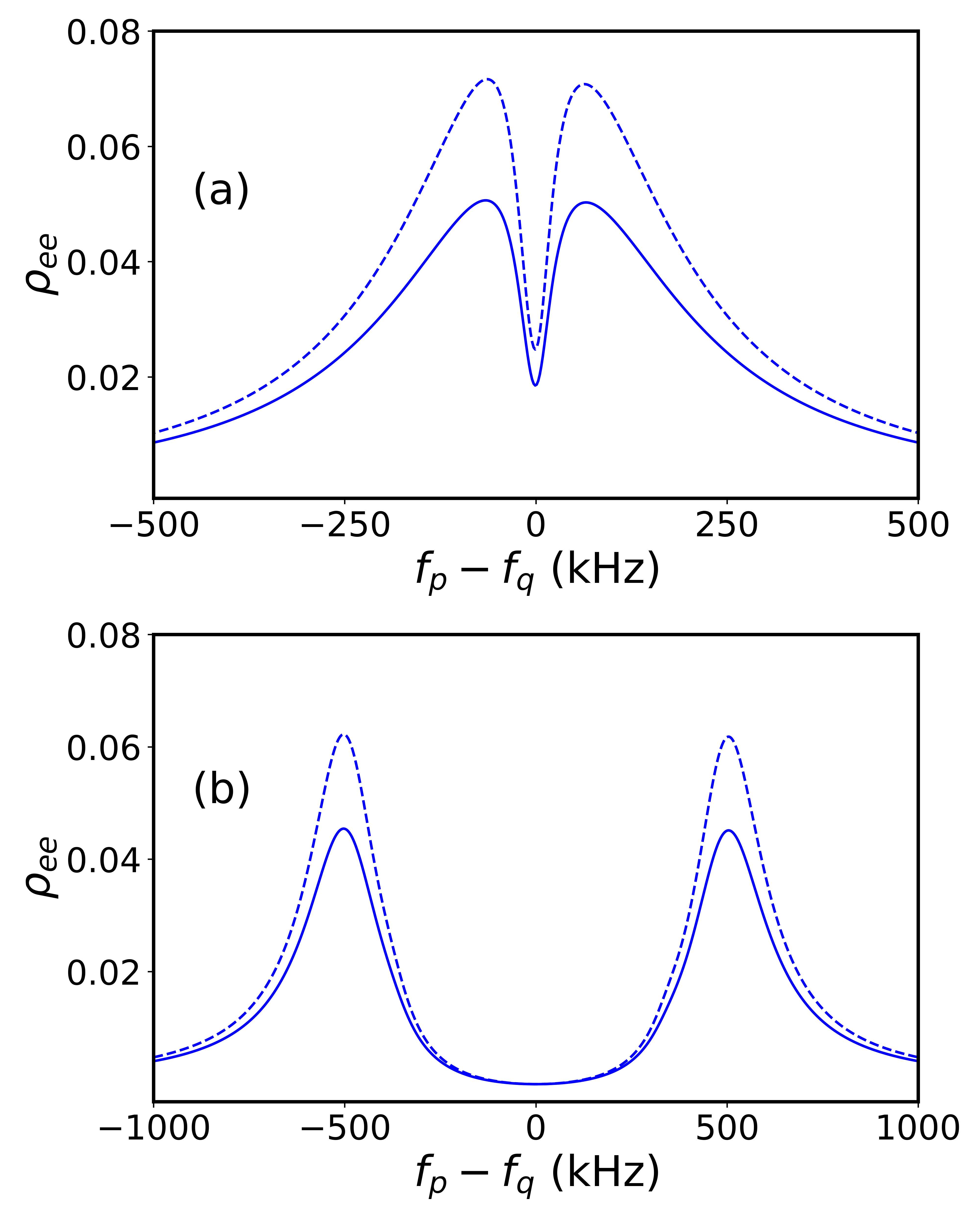}
\caption{Comparison between the qubit population spectrum  with and without considering the pure dephasing rate of the qubit.
(a) Simulated qubit population spectrum in EIT regime when the parameters $\Omega_{sb},\Omega_{p},\gamma,\gamma_{\phi}, \kappa,\chi_{qt}$ are 2$\pi\times$(100,100,400,0,30,10) kHz (solid line) and 2$\pi\times$(100,100,300,50,30,10) kHz (dashed line).
(b) Simulated qubit population spectrum in ATS regime when the parameters $\Omega_{sb}, \Omega_{p},\gamma,\gamma_{\phi},\kappa,\chi_{qt}$ are 2$\pi\times$(1000,100,400,0,30,10) kHz (solid line) and 2$\pi\times$(1000,100,300,50,30,10) kHz (dashed line).}
\label{Pure_dephasing}
\end{figure}

\begin{figure*}[htbp]
\centering\includegraphics[width=1\textwidth]{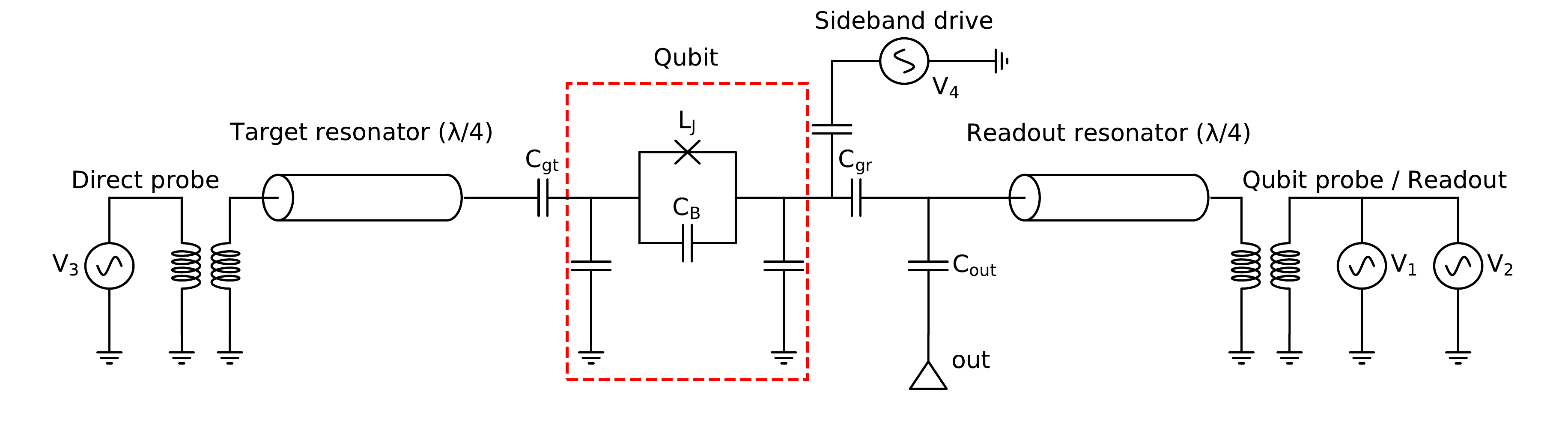}
\caption{A diagram of the device and related electronics used in the experiment.}
\label{circuit_diagram}
\end{figure*}

\begin{figure}[htbp]
\centering\includegraphics[width=0.5\textwidth]{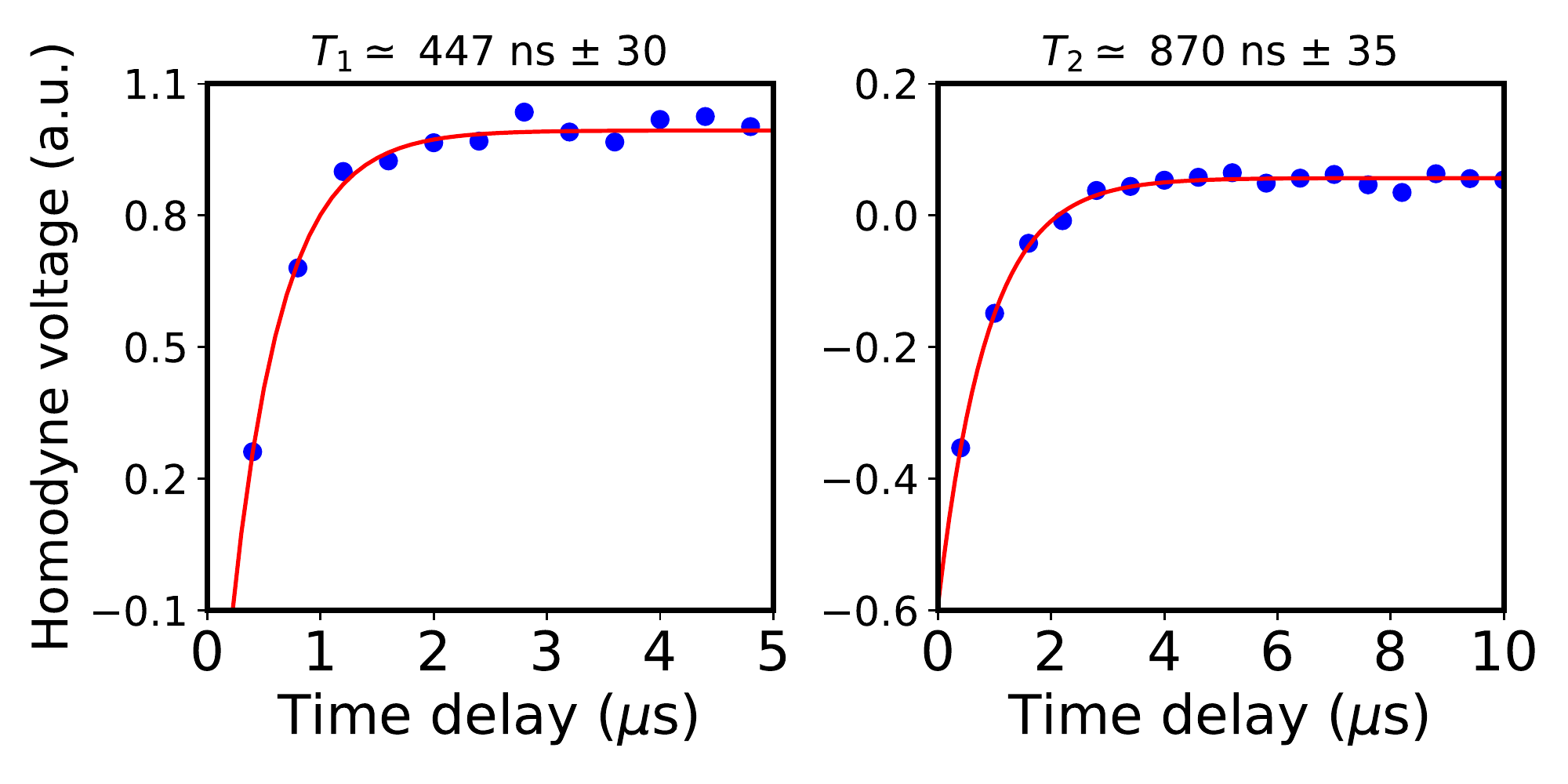}
\caption{$T_{1}$ and $T_{2}$ measurement of the qubit used in the experiment.}
\label{T1T2}
\end{figure}

\begin{figure}[htbp]
\centering\includegraphics[width=0.5\textwidth]{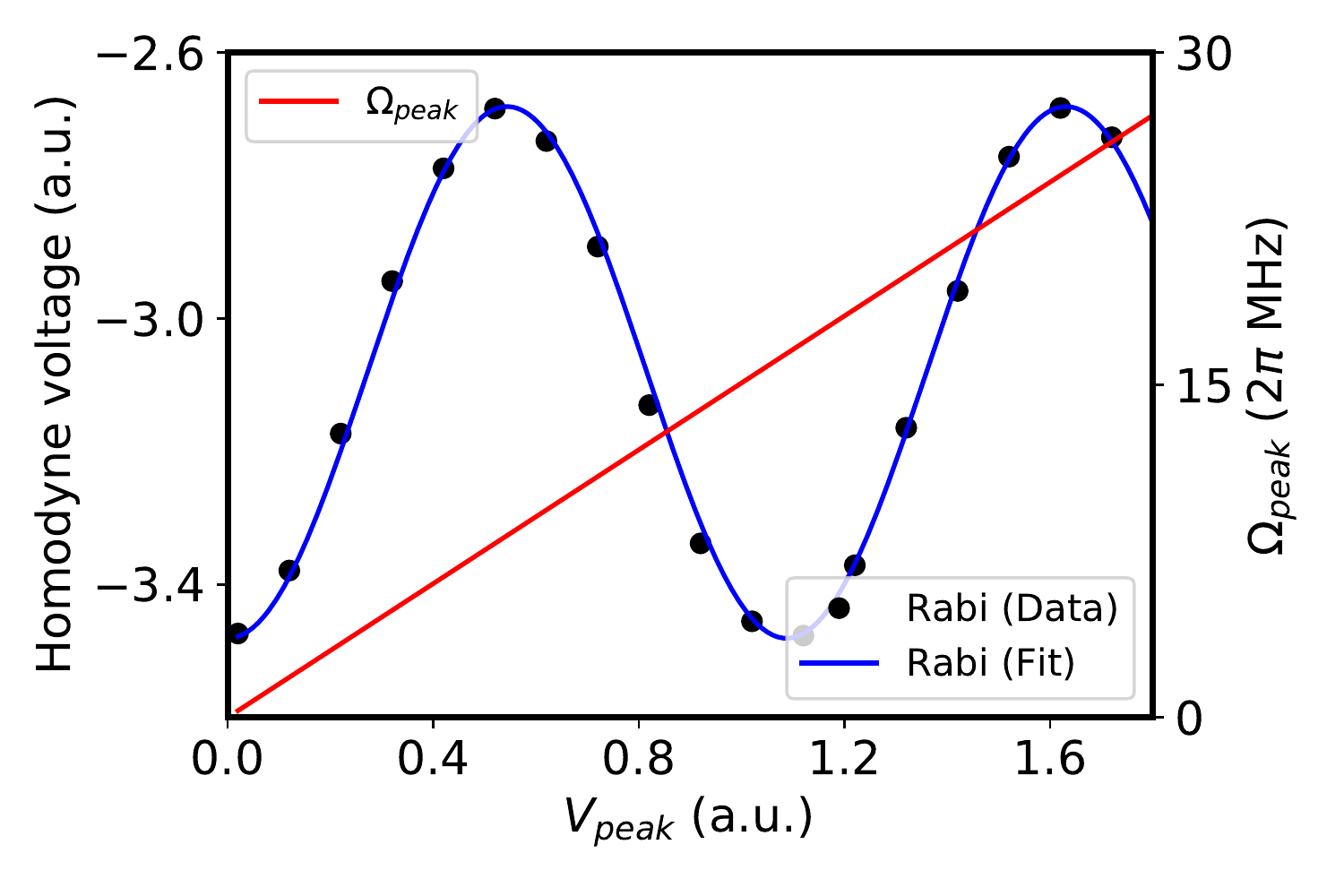}
\caption{Qubit probe amplitude calibration.
A Gaussian Rabi pulse was applied through the readout resonator at the qubit resonant frequency.
The peak voltage of the pulse measured at room temperature is converted to the probe amplitude (red line) in units of angular frequency based on the phase of the oscillation in the homodyne readout signal (black dot and blue line).}
\label{sup1}
\end{figure}

\appendix
\section{Effect of pure dephasing rate in qubit population spectrum}
\label{A0}
In Fig.~\ref{Pure_dephasing},
we present additional simulation results of the qubit population spectrum. For both dashed and solid lines, the qubit have the same total linewidth but different pure dephasing rate, $\gamma_{\phi}$. 
For dashed line,$\gamma_{\phi}$ is $2\pi\times50$kHz whereas for solid line, $\gamma_{\phi}$ is zero. 
One can find that even the total linewidth is the same, we can extract $\gamma_{\phi}$ from the qubit population spectrum.
This is particularly important when using flux-tuneable qubits that normally has significant $\gamma_{\phi}$ comparable to $\gamma$ when the flux is tuned out of the sweep spot. 

\section{Circuit detail}
\label{A1}
In Fig.~\ref{circuit_diagram}, we depict the device and related electronics in the experiment.
A qubit (red dashed box) is coupled to two co-planar waveguide (CPW) resonators. Each resonator is inductively coupled to a separate feedlines. In all, four microwave sources ($V_{1\sim4}$) are used in the experiment: one each for the qubit driving, qubit readout, direct resonator probing, and sideband driving. The device is anchored to the mixing chamber plate of a LD250 Bluefors dilution refrigerator with a base temperature under 7 mK.

\section{Device time domain characteristic}
\label{A2}
In Fig.~\ref{T1T2}, we present a time domain characterization of the qubit used in the experiment.  $T_{1}$ and $T_{2}$ are 447 $\pm$ 30 ns and 870 $\pm$ 35 ns respectively. Since  $2T_{1}\simeq T_2$ holds approximately, it justifies our decision in Section~\ref{2-1} to neglect the pure dephasing in the master equation model.

\section{Probe amplitude calibration}
\label{A3}

We applied a 60-ns long Gaussian pulse with a width of $\sigma$=15 ns at the qubit resonant frequency through the source V1, which was followed by a 200-ns long readout pulse from the same source. The Rabi oscillation swept the peak voltage of the pulse envelop $V_{peak}$ as depicted in Fig.~\ref{sup1}. The phase $\theta$ of this oscillation is given by  $\theta= \Omega_{peak}\int_{-2\sigma}^{2\sigma} \textup{exp}{[-t^{2}/(2\sigma^2)]}dt$. For $\theta=\pi$, $\Omega_{peak}=2\pi\times13.94$ MHz and $V_{peak}=$ 0.54 arb. unit. This yields a conversion factor $\Omega_{peak}/V_{peak}=$25.81 MHz/arb. unit. If the probe field frequency is near the qubit transition frequency, then the probe amplitude $\Omega_p$ is readily calibrated from $V_p$ using this conversion factor. 

\begin{figure}[htbp]
\centering\includegraphics[width=0.5\textwidth]{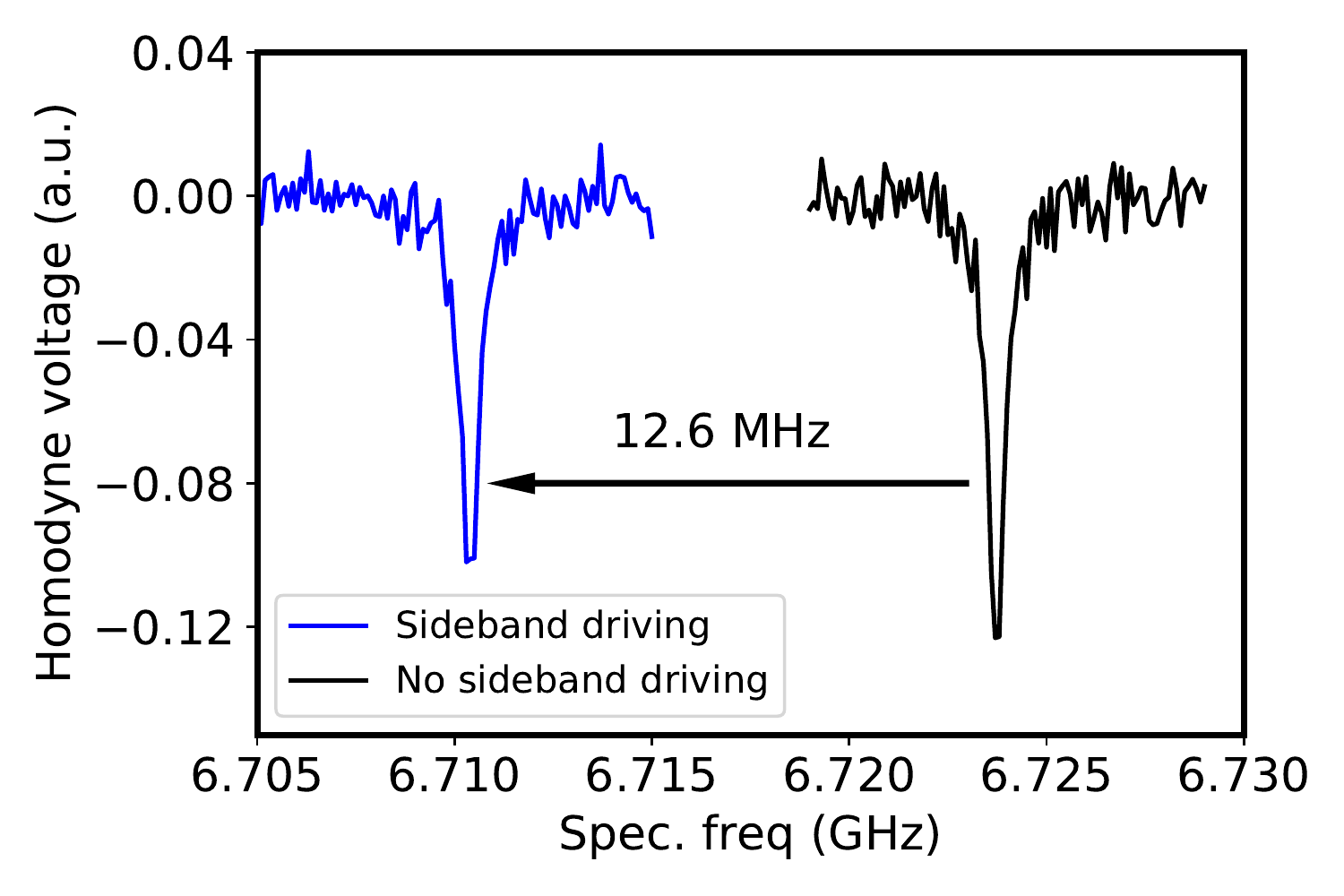}
\caption{Qubit frequency shift under the sideband driving. The amount of the shift is 12.6 MHz when $\Omega_{sb}\approx2\pi \times$ 100 kHz.}
\label{sup2}
\end{figure}

\section{Qubit resonance shift}
\label{A4}

The sideband drive induces not only a sideband transition but also shifts the qubit transition frequency. 
A significant frequency shift (12 MHz downward) is observed when $\Omega_{sb}\approx2\pi\times 100$ kHz.
The data are given in Fig.~\ref{sup2}.


\begin{thebibliography}{10}
\vspace{0.15in}

\bibitem{Bosonic-0} 
M. H. Michael {\em et al.}, \PRX{New Class of Quantum Error-Correcting Codes for a Bosonic Mode
}{6}{031006}{2016}.

\bibitem{Bosonic-01} 
S. Rosenblum {\em et al.}, \Science{Fault-tolerant detection of a quantum error
}{6399}{266}{2018}.

\bibitem{Bosonic-1} 
C. Flühmann {\em et al.}, \Nature{Encoding a qubit in a trapped-ion mechanical oscillator
}{566}{513}{2019}.

\bibitem{Bosonic-2}
R. Lescanne {\em et al.}, {\em Raman coherence in a circuit quantum electrodynamics lambda system
}, Nat. Phys. (2020).

\bibitem{Fano}
A. Miroshnichenko {\em et al.}, {\em Fano Resonances: A Discovery that Was Not Made 100 Years Ago
}, Optics and Photonics News {\bf 19}, 12 (2008).

\bibitem{Fano-2}
M. S. Khalil {\em et al.}, {\em An analysis method for asymmetric resonator transmission applied to
superconducting devices
}, J. Appl. Phys. {\bf 111}, 054510  (2012).

\bibitem{High-Q}
M. Reagor {\em et al.}, {\em Reaching 10ms single photon lifetimes for superconducting aluminum cavities
}, Appl. Phys. Lett. {\bf 102}, 192604 (2013).


\bibitem{eit-1} 
K. M. Birnbaum {\em et al.}, \Nature{Photon blockade in an optical cavity with one trapped atom
}{436}{87}{2005}.

\bibitem{eit-2} 
T. Peyronel {\em et al.}, \Nature{Quantum nonlinear optics with single photons enabled by strongly interacting atoms}{488}{57}{2012}.

\bibitem{eit-3} 
W. Chen {\em et al.}, \Science{All-Optical Switch and Transistor Gated by One Stored Photon
}{6147}{768}{2013}.

\bibitem{Long-PRL2018}
J. Long {\em et al.}, \PRL{Electromagnetically Induced Transparency in Circuit Quantum Electrodynamics with Nested Polariton States}{120}{083602 }{2018}.

\bibitem{eit-4}
G. Andersson {\em et al.}, \PRL{Electromagnetically Induced Acoustic Transparency with a Superconducting Circuit
}{124}{240402 }{2020}.

\bibitem{Novikov-NPhys2016}
S. Novikov {\em et al.}, {\em Raman coherence in a circuit quantum electrodynamics lambda system
}, Nat. Phys. {\bf 12}, 75 (2016).

\bibitem{Liu-PRA2016}
Q. -C. Liu {\em et al.}, \PRA{Method for identifying electromagnetically induced transparency in a tunable circuit quantum electrodynamics system}{93}{053838 }{2016}.

\bibitem{Koch} 
J. Koch {\em et al.}, \PRA{Charge-insensitive qubit design derived from the Cooper pair box
}{76}{042319 }{2007}.

\bibitem{Joo-PRL2010}
J. Joo {\em et al.}, \PRL{Electromagnetically Induced Transparency with Amplification in Superconducting Circuits
}{105}{073601 }{2010}.

\bibitem{Blais}
A. Blais {\em et al.}, \PRA{Quantum information processing with circuit quantum electrodynamics
}{75}{032329}{2007}.

\bibitem{Gu-PRA2016}
X. Gu {\em et al.}, \PRA{Polariton states in circuit QED for electromagnetically induced transparency
}{93}{063827 }{2016}.

\bibitem{PNS}
D. I. Schuster {\em et al.}, \Nature{Resolving photon number states in a superconducting circuit
}{445}{515)}{2007}.

\bibitem{Mika-PRL2009}
M. A. Sillanpää {\em et al.}, \PRL{Autler-Townes Effect in a Superconducting Three-Level System
}{103}{193601)}{2009}.

\bibitem{Kelly-PRL2010}
W. R. Kelly {\em et al.}, \PRL{Direct Observation of Coherent Population Trapping in a Superconducting Artificial Atom
}{104}{163601}{2010}.

\bibitem{Li-SciRep}
J. Li {\em et al.}, {\em Dynamical Autler-Townes control of a phase qubit
}, Sci. Rep. {\bf 2}, 645   (2012).

\bibitem{Novikov-PRB2013}
S. Novikov {\em et al.}, \PRB{Autler-Townes splitting in a three-dimensional transmon superconducting qubit
}{88}{060503(R)}{2013}.

\bibitem{SRON}
D. J. Thoen {\em et al.}, {\em Superconducting NbTiN Thin Films with Highly Uniform Properties Over a Ø 100 mm Wafer}, IEEE Transactions on Applied Superconductivity, {\bf 27}, (2017).

\bibitem{Anisimov-PRL2011}
P. M. Anisimov {\em et al.}, \PRL{Objectively Discerning Autler-Townes Splitting from Electromagnetically Induced Transparency
}{107}{163604 }{2011}.

\bibitem{Klimov-PRL2018}
P. V. Klimov {\em et al.}, \PRL{Fluctuations of Energy-Relaxation Times in Superconducting Qubits
}{121}{090502}{2018}.

\bibitem{Xmon}
Y. Chen {\em et al.}, \PRL{Qubit architecture with high coherence and fast tunable coupling
}{113}{220502}{2014}.

\bibitem{Starmon}
R. Versluis {\em et al.}, \PRApplied{Scalable Quantum Circuit and Control for a Superconducting Surface Code
}{8}{034021}{2017}.


\bibitem{data}
10.5281/zenodo.3956080.

\end{thebibliography}
\end{document}